\documentclass[fleqn,usenatbib]{mnras}

\usepackage{newtxtext,newtxmath}

\usepackage[T1]{fontenc}

\DeclareRobustCommand{\VAN}[3]{#2}
\let\VANthebibliography\thebibliography
\def\thebibliography{\DeclareRobustCommand{\VAN}[3]{##3}\VANthebibliography}

\usepackage{amsmath}
\usepackage{wasysym}
\usepackage[inter-unit-product=\cdot]{siunitx}
\usepackage{graphicx}
\usepackage{cleveref}

\newcommand{\V}[1]{\boldsymbol{#1}}
\newcommand{\dd}[3][]{\dfrac{\mathrm{d}^{#1}{#2}}{\mathrm{d}{#3}^{#1}}}
\newcommand{\pp}[3][]{\dfrac{\partial^{#1}{#2}}{\partial{#3}^{#1}}}
\newcommand{\dif}{{\mathrm d}}
\newcommand{\lmean}[1]{\left\langle {#1} \right\rangle}
\newcommand{\msun}{M_{\astrosun}}
\newcommand{\tconv}{\tau_{\text{conv}}}
\newcommand{\rstar}{R_{\star}}

\title[Effects of stratification in evolved stars]{Effects of stratification on overshooting and waves atop the convective core of $5\msun$ main-sequence stars}

\author[A. Morison et al.]{
A. Morison,$^{1}$\thanks{E-mail: a.morison@exeter.ac.uk}
A. Le Saux,$^{2}$
I. Baraffe,$^{1,3}$
J. Morton,$^{1}$
T. Guillet,$^{1}$
D. G. Vlaykov,$^{4}$
T. Goffrey,$^{5,6}$
and J. Pratt$^{7}$
\\
$^{1}$Physics and Astronomy, University of Exeter, Exeter EX4 4QL, UK\\
$^{2}$Laboratoire de Météorologie Dynamique (IPSL), Sorbonne Université, CNRS, Ecole Polytechnique, Ecole Normale Supérieure, Paris, France\\
$^{3}$Ecole Normale Supérieure de Lyon, CRAL (UMR CNRS 5574), Université de Lyon, 69007 Lyon, France\\
$^{4}$Mathematics and Statistics, University of Exeter, Exeter EX4 4QF, UK\\
$^{5}$Centre for Fusion, Space and Astrophysics, Department of Physics, University of Warwick, Coventry CV4 7AL, UK\\
$^{6}$Healthcare Technology Institute, School of Chemical Engineering, The University of Birmingham, Birmingham, B15 2TT, UK\\
$^{7}$Lawrence Livermore National Laboratory, 7000 East Ave, Livermore, CA 94550, USA
}

\date{Accepted XXX. Received YYY; in original form ZZZ}

\pubyear{2024}

\begin{document}
\label{firstpage}
\pagerange{\pageref{firstpage}--\pageref{lastpage}}
\maketitle

\begin{abstract}
As a massive star evolves along the main sequence, its core contracts, leaving
behind a stable stratification in helium. We simulate 2D convection in the
core at three different stages of evolution of a $5\msun$ star, with three
different stratifications in helium atop the core. We study the propagation
of internal gravity waves in the stably-stratified envelope, along with the
overshooting length
of convective plumes above the convective boundary. We find
that the stratification in helium in evolved stars hinders radial motions
and effectively shields the radiative envelope against plume penetration.
This prevents convective overshooting from being an efficient mixing
process in the radiative envelope.
In addition, internal gravity waves are less excited in evolved models compared
to the zero-age-main-sequence model, and are also more damped in the stratified
region above the core.
As a result, the wave power is several orders of magnitude lower in mid-
and terminal-main-sequence models compared to zero-age-main-sequence stars.
\end{abstract}




\section{Introduction}
\label{sec:intro}

One-dimensional formalisms for mixing due to convective overshooting and
internal gravity waves have the potential to improve the comparison between
stellar evolution models and observations. Several mechanisms
have been proposed to contribute to the radial transport of chemical elements: mainly
overshooting of plumes from a convective region in a surrounding radiative
region \citep{shaviv_convective_1973,stancliffe_confronting_2016},
transport via internal waves \citep{schatzman_transport_1993}, or erosion of compositional
gradient via shearing \citep[possibly enhanced by rotation,][]{zahn_circulation_1992}.
The mixing induced by these mechanisms is however typically too weak to convincingly
explain several observations. Such observations include evolutionary tracks of eclipsing binaries
\citep{claret_dependence_2016}, constraints from color-magnitude diagrams
of stellar clusters in the Large Magellanic Cloud \citep{rosenfield_new_2017}, and
constraints from asteroseismology \citep{bossini_uncertainties_2015}.
Recently, \citet{baraffe_study_2023}
showed that convective overshooting above the cores of massive stars determined
from 2D hydrodynamical simulations is too
weak to explain the observed width of the main sequence in the Hertzsprung–Russell
diagram. Convective
overshooting was measured in simulations at the Zero-Age Main Sequence (ZAMS).
The present study is a follow-up of the latter, where we also measure
the extent of
convective overshooting at later times along the main sequence: mid main
sequence (MidMS), and terminal main-sequence (TAMS). This study provides hints
at how convective overshooting changes as stars evolve along the main sequence
and whether this mechanism is a viable candidate for radial mixing.

Another open question in the field of stellar dynamics is that of the
propagation of internal gravity waves (IGW) generated at the convective core
boundary and whether they can be observed \citep{bowman_photometric_2019,
bowman_photometric_2020, lecoanet_low-frequency_2019, lecoanet_surface_2021}.
Recently
\cite{vanon_three-dimensional_2023} showed simulations of evolved $7\msun$ stars
on the main sequence
where the propagation of IGW is greatly affected by the Brunt-Väisälä peak
above the convective core. \cite{ratnasingam_internal_2023} showed simulations
of $3-13\msun$ at a mid-main-sequence age, also pointing out the effects of the
Brunt-Väisälä peak on the propagation of waves. We perform a similar study for
our three models of $5\msun$ stars along the main sequence.
Two key differences of our simulations compared to previous ones is that ours
are fully compressible instead of anelastic, and use temperature and helium
mass fractions profiles from stellar evolution models instead of prescribing
the $N^2$-peak region via a modified temperature profile.

In \Cref{sec:methods}, we detail the models corresponding to the three
evolutionary stages along the main sequence. We present our setup for the
2D numerical simulations with the fully compressible time-implicit code MUSIC,
and how we assess the penetration
length above the convective core in those simulations. We then present
the theoretical framework we use to study internal gravity waves that
propagate in the radiative region. \Cref{sec:results} groups our
results, showing the consequences of the helium stratification on the
dynamics, penetration length, and IGW propagation. We end this paper
with discussions and conclusive remarks in \Cref{sec:conclusion}.

\section{Methods}
\label{sec:methods}

We focus on three evolutionary stages of a $5\msun$ star along the main
sequence: zero age (ZAMS), mid main sequence (MidMS) and terminal main sequence
(TAMS).  All three models are
part of the same evolutionary path of a $5\msun$-star, with an initial helium
mass fraction $Y=0.28$ and metallicity $Z=0.02$. \Cref{tab:cases} summarizes the
three cases.

\subsection{1D stellar models}

We use the one-dimensional (1D) stellar evolution code presented in
\cite{baraffe_evolution_1991,baraffe_evolutionary_1998} to compute 1D models
for each of those three stages.

\begin{table}
    \begin{center}
    \begin{tabular}{*{8}c}
        model & $\rstar$ (\unit{\cm}) & $R_\text{conv}/\rstar$ & $H_{p,\text{conv}}/\rstar$ & $L/L_{\odot}$ & $Y_\text{core}$ \\
        \hline
        ZAMS & \num{1.84e11} & 0.18  &  0.099 & 522 & 0.28 \\
        MidMS & \num{2.51e11} & 0.12  &  0.070 & 680 & 0.56 \\
        TAMS & \num{2.92e11} & 0.095 &  0.056 & 758 & 0.70 \\
    \end{tabular}
    \end{center}
    \caption{Physical parameters of the three evolutionary stages. $\rstar$ is
    the total radius of the star, $R_\text{conv}$ is the position of the
    Schwarzschild boundary of the convective core, $H_{p,\text{conv}}$ is the
    pressure scale height at that boundary, $L$ is the total luminosity of
    the star, and $Y_\text{core}$ is the helium mass fraction in the core.}
    \label{tab:cases}
\end{table}

\begin{figure}
\begin{center}
    \includegraphics[width=\columnwidth]{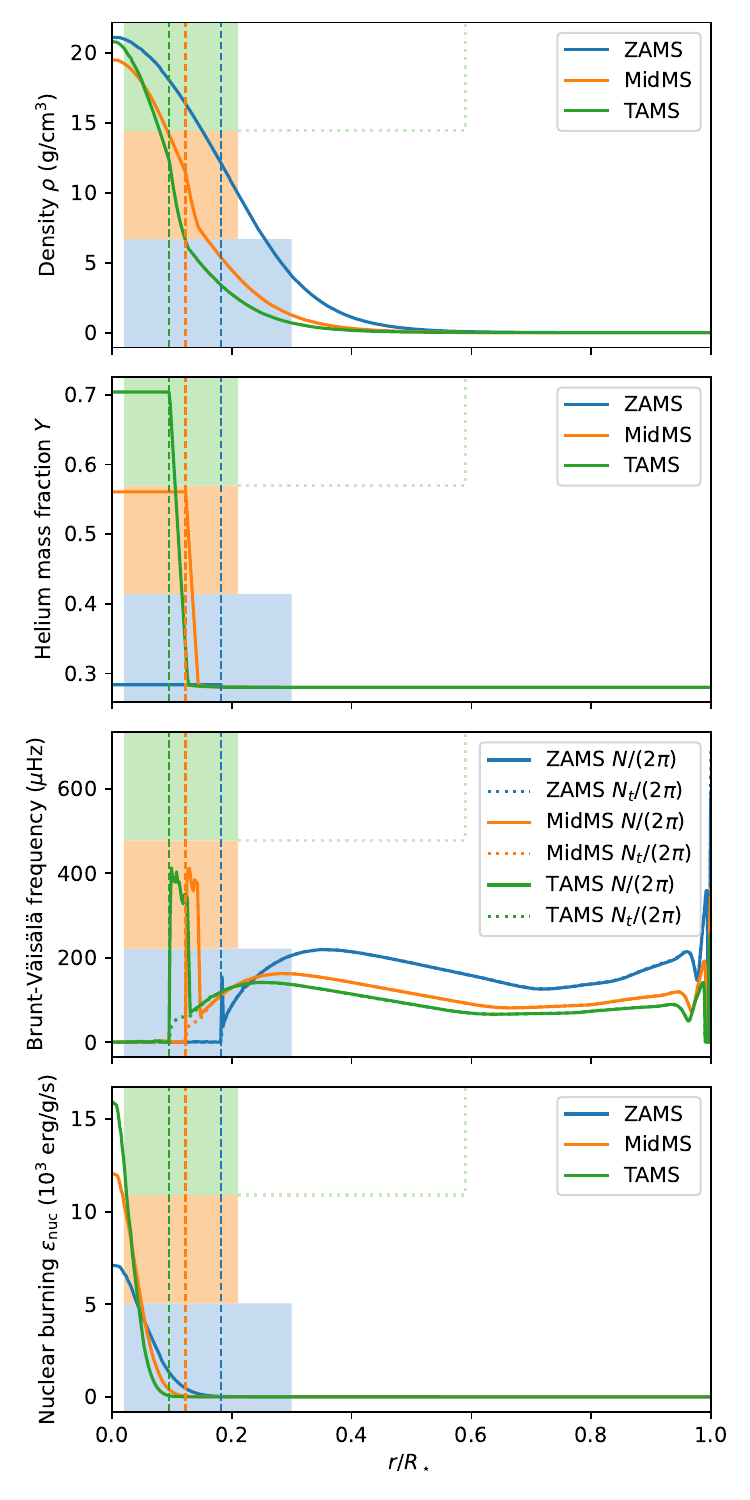}
\end{center}
\caption{Radial profiles for the three stages of evolution from the 1D
    evolution model. Those profiles are then interpolated to use them as
    initial condition for the 2D simulations. Dashed lines mark the position of
    the convective core boundary for each model according to the Schwarzschild
    criterion. Shaded areas mark the radial extent of the simulated domain for
    each evolution stage. The dotted box shows the additional radial domain in
    the ``TAMS ext'' model (identical to the TAMS model but with a higher
    external radius).
    See \cref{eq:brunt} for the definition of $N$ and
    \cref{eq:nthermal} for the definition of $N_t$.}
\label{fig:initprof}
\end{figure}

\Cref{fig:initprof} shows profiles from these three 1D models. One interesting feature
is the helium mass fraction profiles that vary widely above the convective core
across the three evolution stages considered
here. The core retracts and its helium content increases
as a result of hydrogen burning as the star evolves along the main sequence. The convective
core is assumed to be well-mixed and therefore to have a radially constant helium fraction, and most
of the stable envelope stays at the initial helium mass fraction $Y=0.28$. Due to the core
retracting, a layer therefore develops atop of the convective core with a strong stratification
in helium. This gradient of concentration in helium directly translates into a strong
density gradient above the convective core, and therefore a region where the
Brunt-Väisälä frequency is much higher than in the rest of the radiative
envelope (around \SI{400}{\micro\Hz} versus \SI{100}{\micro\Hz}). This region
is hence referred to as the $N^2$-peak region in this paper.

The 1D stellar models used in this study rely on the Schwarzschild criterion for the onset of
convective instability and do not account for overshooting at the convective
core boundary. Because the convective core retracts during core H-burning, the
presence of this $N^2$-peak layer and of a molecular weight gradient above the
convective core is robust. Assumptions used in the 1D models can certainly
affect the properties of this layer (width, radial profile of the molecular
weight gradient), but not its existence. \citet{miglio_probing_2008} find for
instance that accounting for convective overshooting in a $6\msun$ model only
marginally increases the radial extent of the $N^2$-peak layer but not its
amplitude; they report however more drastic differences for $1.6\msun$ stars.
Therefore we expect that for $5\msun$ stars, the impacts of the
$N^2$-peak layer on convective penetration and waves that we report in this
work will remain qualitatively the same independently of the assumptions used
to construct the 1D models.

\subsection{2D numerical simulations}

The 1D profiles described previously are fed as initial condition for the fully
compressible stellar hydrodynamics code MUSIC
\citep{viallet_jacobian-free_2016, goffrey_benchmarking_2017}. We solve the
mass, momentum, internal energy, and helium conservation equations for a fully
compressible inviscid fluid:
\begin{equation}
    \label{eq:mass}
    \pp{\rho}{t} = -\nabla\cdot (\rho\V v),
\end{equation}
\begin{equation}
    \label{eq:momentum}
    \pp{\rho \V v}{t} = -\nabla\cdot (\rho\V v\otimes \V v) -\nabla p + \rho \V g,
\end{equation}
\begin{equation}
    \label{eq:energy}
    \pp{\rho e}{t} = -\nabla\cdot (\rho e\V v) - p\nabla\cdot \V v + \nabla\cdot (\chi\nabla T) + \rho \varepsilon_\text{nuc},
\end{equation}
\begin{equation}
    \label{eq:helium}
    \pp{\rho Y}{t} = -\nabla\cdot (\rho Y\V v),
\end{equation}
where $\rho$ is the fluid density, $e$ the specific internal energy, $\V v$ the
fluid velocity, and $Y$ the helium mass fraction.
The gravitational acceleration $\V g=-g(r)\hat{\V r}$ is purely
radial and updated after each time step from the instantaneous density profile
$\bar\rho(r)$:
\begin{equation}
    g(r) =  4\pi\frac{G}{r^2} \int_0^r \bar\rho(u)u^2du.
\end{equation}
The density profile $\bar\rho$ is calculated as
\begin{equation}
    \label{eq:densprof}
    \bar\rho = \lmean{\rho}_\theta
\end{equation}
where $\lmean{\bullet}_\theta$ denotes an angular average across the domain
accounting for sphericity
\begin{equation}
    \label{eq:meantheta}
    \lmean{f}_\theta = \frac12\int_0^\pi f\sin\theta\dif\theta.
\end{equation}
We recover the pressure $p$ and temperature $T$ from $\rho$ and $e$ using
tabulated equation of state, which for the simulated regime of $5\msun$
main-sequence stars are the OPAL tables from \citet{rogers_updated_2002}.
Radiative transfer is accounted for via
a diffusive formulation involving the heat conductivity $\chi$. The latter is derived
from the Rosseland mean opacity $\kappa$ given by the OPAL opacity tables
\citep{iglesias_updated_1996}:
\begin{equation}
    \label{eq:rad_cond}
    \chi = \frac{16\sigma T^3}{3\kappa \rho}
\end{equation}
(with $\sigma$ the Stefan-Boltzmann constant).
Note that the
1D stellar evolution code and MUSIC use the same equation of state tables and
opacities for consistency.
Finally, $\varepsilon_\text{nuc}$ is the specific energy released by nuclear
burning. Its radial profile is taken from the 1D model (see \Cref{fig:initprof}),
and considered constant in time throughout the 2D run since
nuclear burning evolves on much longer timescales than the duration of the
simulations. This is also why no variation of $Y$ due to nuclear reactions appears
in \cref{eq:helium}.

All the simulations performed in this study are 2D in $(r,\theta)$ space, assuming
$v_\phi=0$ and invariance along $\phi$ (see \Cref{tab:sims} and hereafter for details).
For all three cases, we solve in a domain
that spans the entire latitudinal range ($\theta\in[0, \pi]$), and truncate
along the radial direction ($r\in[R_\text{in}, R_\text{out}]$, see \Cref{tab:sims}). All simulations are performed
on a uniform mesh
at a similar resolution per pressure scale height at the Schwarzschild
boundary (denoted $R_\text{conv}$): $H_{p,\text{conv}}\sim 140\Delta r$, with $\Delta r$ the grid size
in radial direction. With this resolution, the jump in helium mass fraction
above the core spans across $\sim40\Delta r$ in the MidMS model and
$\sim80\Delta r$ in the TAMS model.
We use reflective boundary conditions at $\theta=0$ and $\theta=\pi$. At $R_\text{in}$
and $R_\text{out}$, we impose reflective boundary conditions on velocity,
a constant radial derivative in density \citep[following e.g.][]{baraffe_study_2023}, and the energy
fluxes taken from the 1D model at those radii.

MUSIC uses a fully implicit time scheme \citep{viallet_jacobian-free_2016};
timesteps are therefore not limited by stability considerations. The timestep
is set based on limits on both the acoustic and advective CFL. We pick the
maximum timestep $\Delta t$ that satisfies both $\Delta t\frac{c_S +
|v_i|}{\Delta r} \leqslant 100$ and $\Delta t\frac{|v_i|}{\Delta r} \leqslant
0.1$ for every velocity component $v_i$ everywhere in the simulation domain,
where $c_S$ is the local sound speed. This corresponds to an acoustic and
advective Courant number of 100 and 0.1, respectively. In practice, this leads
to timesteps of the order of a minute.

As the initial conditions for the 2D solver are built from 1D models, the 2D
models go through an initial transient as the 2D convective motions and waves
develop.  In this study, we focus on the statistical steady state reached by
the models after this transient stage.

To make sure our simulation time in steady state is long enough to capture
meaningful statistics, we use the convective timescale $\tconv$ defined as
\begin{equation}
    \label{eq:tconv}
    \tconv = \lmean{\int_{R_\text{in}}^{R_\text{conv}}\frac{\dif r}{v_\text{rms}(r,t)}}_t
\end{equation}
where $\lmean{\bullet}_t$ denotes a time average across the steady state.
$v_\text{rms}$ is the instantaneous root-mean-square velocity profile:
\begin{equation}
    \label{eq:vrms}
    v_\text{rms} = \sqrt{\lmean{\V v\cdot \V v}_\theta}.
\end{equation}
The models being 2D and of relatively modest resolution allows us to run the
simulations for a long time, achieving about a thousand of convective times for
all three models (see \Cref{tab:sims}). Such simulation times are necessary for
the diagnostics related to overshooting length based on extreme events to
converge \citep{pratt_extreme_2017, baraffe_study_2023}. A quantity associated
with the convective timescale is the convective frequency
\begin{equation}
    \label{eq:freqconv}
    f_\text{conv} = \frac{1}{\tconv}.
\end{equation}
Notice on \Cref{tab:sims} that the convective frequency is higher in models
that are further along on the main sequence (due to higher convective
velocities), leading to higher frequency waves being excited.
``TAMS ext'' in \Cref{tab:sims} is a check for the impact of the low
cut-off radius on internal gravity waves in evolved models: it is identical
to the ``TAMS'' model but with a larger external radius, to verify that the
features observed with the small cut-off radius are also observed in the
larger domain.

\begin{table*}
    \begin{tabular}{*{9}c}
        model & $\tconv$ (s) & $f_\text{conv}$ (\unit{\micro\Hz}) & $R_\text{in}/\rstar$ & $R_\text{conv}/\rstar$ & $R_\text{out}/\rstar$ & $\tau_\text{steady}/\tconv$ & $n_r\times n_\theta$ & $H_{p,\text{conv}}/\Delta r$ \\
        \hline
        ZAMS & \num{1.6e6} & 0.625 & 0.02 & 0.18 & 0.3 & 1188 & $400\times 200$ & 141\\  
        MidMS & \num{8.7e5} & 1.15 & 0.02 & 0.12 & 0.21 & 1368 & $384\times 192$ & 142\\  
        TAMS & \num{6.7e5} & 1.49 & 0.02 & 0.095 & 0.21 & 910 & $472\times 236$ & 140\\  
        TAMS ext & \num{6.4e5} & 1.56 & 0.02 & 0.095 & 0.59 & 31 & $1416\times 236$ & 140\\  
    \end{tabular}
    \caption{Setup for 2D simulations of the three evolution stages. ``TAMS
    ext'' is an extended simulation that uses the same initial condition as
    the TAMS model, but with a higher external radius. This is to check
    the impact of the fairly low cut-off radius on internal gravity waves in the
    other models. $\tconv$ is the convective timescale \cref{eq:tconv},
    $f_\text{conv}$ the convective frequency \cref{eq:freqconv},
    $R_\text{in}$ the inner radius of the simulated domain, $R_\text{out}$ the
    outer radius of the simulated domain, $R_\text{conv}$ the Schwarzschild
    radius of the convective core, $\tau_\text{steady}$ the duration of the steady-state
    of the simulation,
    $n_r, n_\theta$ the number of grid cells along the radial and latitudinal directions,
    $H_{p,\text{conv}}$ the pressure scale height at $R_\text{conv}$, $\rstar$
    the radius of the star, and $\Delta r$ the grid cell size along the radial
    direction.}
    \label{tab:sims}
\end{table*}

\subsection{Penetration length}
\label{sub:pendepth}
A first aspect that this study focuses on is the impact of the
stratification in helium atop the core on overshooting length.
This follows the method developed in \citet{pratt_extreme_2017} and used in
\citet{baraffe_two-dimensional_2021,baraffe_study_2023},
of which an overview is done here.
This method is built around two different fluxes:
the radial kinetic energy flux $f_k$
\begin{equation}
    \label{eq:ekinflux}
    f_k(r,\theta,t) = \frac 12\rho\V v^2v_r
\end{equation}
and the radial heat flux $f_{\delta T}$
\begin{equation}
    \label{eq:thflux}
    f_{\delta T}(r,\theta,t) = \rho c_p \delta T v_r.
\end{equation}
The temperature fluctuations $\delta T$ are the departure from the mean
temperature profile of the steady state:
\begin{equation}
    \delta T = T - \lmean{\lmean{T}_\theta}_t
\end{equation}
The overshooting distance $l_k$ with respect to $f_k$ is then defined as
\begin{equation}
    \label{eq:lkin}
    l_k(\theta, t) = r_k - R_\text{conv}
\end{equation}
where $r_k$ is the first radius above $R_\text{conv}$ at which $f_k=0$.
Similarly, $l_{\delta T}(\theta,t)$ is defined from the zero-crossing
of $f_{\delta T}$. For such low mass models, we do not observe any significant
penetration (significant modification of the temperature gradient above the
convective core was found in \citet{baraffe_study_2023} for more luminous
models); the position of the convective boundary $R_\text{conv}$ in
\cref{eq:lkin} is therefore taken from the 1D model (see \Cref{tab:cases}).
Following \citet{pratt_extreme_2017}, we assess the overshooting length via the
angular maximum of $l_k(\theta, t)$:
\begin{equation}
    \label{eq:lmaxt}
    l_k^\text{max}(t) = \max_\theta l_k(\theta, t)
\end{equation}
and, following \citet{baraffe_study_2023}, its time average defines the
overshooting length
\begin{equation}
    \label{eq:lmax}
    l_k^\text{ov}= \lmean{l_k^\text{max}}_t
\end{equation}
and similarly for $l_{\delta T}^\text{max}$ and $l_{\delta T}^\text{ov}$.
Several hundreds of overturn times (see \Cref{tab:sims}) are needed for
$l_k^\text{ov}$ and $l_{\delta T}^\text{ov}$ to converge to a steady
value. Once that steady state reached, both diagnostics give a similar
value for the overshooting length, as discussed in
\cite{pratt_extreme_2017,baraffe_study_2023}.

\subsection{Waves study}

The second aspect on which this study focuses is the impact of the stably
stratified layer atop of the core on the excitation and propagation of internal
gravity waves in the envelope. This follows the work presented in
\citet{le_saux_two-dimensional_2022, lesaux_two-dimensional_2023}.
The reader is referred to those studies for more details, an overview of
the method is presented here for convenience.

In a stably-stratified fluid, a particle of fluid that is out-of-equilibrium
(i.e. denser or lighter than its surroundings) is brought back to its
equilibrium position by the buoyancy force. This mechanism allows IGW to
propagate in a stably-stratified fluid. The angular Brunt-Väisälä frequency $N$
(in \unit{\radian/\s})
is the maximum frequency of such waves. It is directly related to the
density stratification of the medium:
\begin{equation}
    \label{eq:brunt}
    N^2 = g\left(\frac{1}{\Gamma_1}\dd{\ln p}{r} - \dd{\ln \rho}{r}\right).
\end{equation}
$\Gamma_1$ is the first adiabatic exponent:
\begin{equation}
    \label{eq:gamma1}
    \Gamma_1 = \left.\pp{\ln p}{\ln \rho}\right|_\text{ad}.
\end{equation}
A fluid is unstable against convection if $N^2\leqslant 0$, in which case IGW are
evanescent. Conversely, a fluid is stably-stratified if $N^2>0$, and IGW of
angular frequency $\omega \leqslant N$ can propagate in the medium.
Note that in this paper, $\omega$ denotes the angular frequency in \unit{\radian/\s},
and $f=\omega/(2\pi)$ is the frequency in hertz.
In the absence of
rotation, the dispersion relation of IGW verifies \citep{press_radiative_1981}:
\begin{equation}
    \label{eq:dispersion}
    \frac{\omega ^2}{N^2} = \frac{k_h^2}{k_r^2 + k_h^2}
\end{equation}
where $k_r$ is the radial wavenumber and $k_h$ the horizontal wavenumber. The
latter is related to the spherical harmonic degree $\ell$ of the considered wave:
\begin{equation}
    \label{eq:kh}
    k_h^2 = \frac{\ell(\ell + 1)}{r^2}.
\end{equation}

The linear theory developed in \citet{press_radiative_1981} and extended to
variable molecular weight $\mu$ in \citet{zahn_angular_1997} uses the WKB
approximation to predict the amplitude of travelling IGW in the radiative
region. This yields the following expression for the radial velocity:
\begin{equation}
    \label{eq:linampl}
    \hat v_r(r,\ell,\omega) = \hat v_e(\ell,\omega)
    \left(\frac{\rho_e}\rho\right)^{1/2}
    \left(\frac{r_e}{r}\right)^{3/2}
    \left(\frac{N_e^2-\omega^2}{N^2-\omega^2}\right)^{1/4}
    \exp(-\tau/2)
\end{equation}
where $e$-indices denote quantities evaluated at the excitation radius. Note
that $\hat v_e(\ell,\omega)$ is an arbitrary amplitude for the linear theory.
In \cref{eq:linampl}, $\tau$ represents the damping of
waves due to radiative effects. Radiative diffusion is the main damping
mechanism of IGW in stellar interiors, and the only one considered in this
linear model.
$\tau$ verifies \citep{zahn_angular_1997}:
\begin{equation}
    \label{eq:lindamping}
    \tau(r,\ell,\omega) =
    \int_{r_e}^r\kappa_{\text{rad}} k_h^3 \frac{NN_t^2}{\omega^4}
    \left(\frac{N^2}{N^2-\omega^2}\right)^{1/2}
    dr.
\end{equation}
$\kappa_\text{rad}$ is the radiative diffusivity
\begin{equation}
    \label{eq:krad}
    \kappa_\text{rad} = \frac{\chi}{\rho c_p}
\end{equation}
where $c_p$ is the specific heat capacity as constant pressure.
$N_t$ is the thermal contribution to the angular Brunt-Väisälä frequency
\begin{equation}
    \label{eq:nthermal}
    N_t^2 = -\left.\pp{\ln\rho}{\ln T}\right|_{p,\mu}
    \cdot \frac{g}{H_p} \cdot
    \left(\left.\pp{\ln T}{\ln p}\right|_\text{ad} - \dd{\ln T}{\ln p}\right).
\end{equation}
A related quantity is the wave luminosity, defined as
\citep{press_radiative_1981, zahn_angular_1997}:
\begin{equation}
    \label{eq:lwave}
    L_{\text{wave}} = 4\pi r^2 \rho v_\text{rms}^2 V_{g,r}
\end{equation}
where $V_{g,r}$ is the radial group velocity
\begin{equation}
    \label{eq:vgroup}
    V_{g,r} = \frac{\omega^2}{N^2}\frac{\sqrt{N^2-\omega^2}}{k_h}.
\end{equation}
This luminosity is only affected by damping as waves travel in the radiative
region:
\begin{equation}
    \label{eq:lwavelin}
    L_\text{wave}(r,\ell,\omega) = L_{\text{wave}}(r_e,\ell,\omega)\exp(-\tau).
\end{equation}

To identify and study IGW in the radiative envelope of our simulations, we
use a temporal Fourier transform and decomposition in spherical harmonics
of the radial velocity field of the 2D models. $P[\hat v_r](r,\ell,f)$
denotes the power spectrum obtained from such a spectral analysis \citep[see
Appendix A of][for more details]{le_saux_two-dimensional_2022}. This power
spectrum is directly the rms of the radial velocity. Using $v_h/v_r\simeq k_r/k_h$
and the dispersion relation \cref{eq:dispersion}, the wave luminosity \cref{eq:lwave}
can then be expressed from this power spectrum:
\begin{equation}
    \label{eq:lwavespec}
    L_\text{wave} \simeq 4\pi r^2 \rho \frac{\sqrt{N^2 - (2\pi f)^2}}{k_h} P[\hat v_r].
\end{equation}

The linear expression on $\hat v_r$ \cref{eq:linampl} also applies to
$\sqrt{P[\hat v_r]}$ since $\hat v_r=\sqrt{P[\hat v_r]/2}$. For simplicity, we
test the linear prediction on profiles of $\sqrt{P[\hat v_r]}$ directly.
The excitation radius $r_e$ in \cref{eq:linampl} is taken slightly above the
convective core boundary, at twice the overshooting length \cref{eq:lmax}:
\begin{equation}
    \label{eq:re}
    r_e = R_\text{conv} + 2 l_k^\text{ov}.
\end{equation}
Note that taking $r_e$ at $l_k^\text{ov}$ or $2l_k^\text{ov}$ above the
convective boundary does not change our results significantly. To
improve the readability of figures, we choose $\hat v_e(\ell,\omega)$
in \cref{eq:linampl}
as the local maximum of $\sqrt{P[\hat v_r]}$ evaluated around $r_e$.
Comparing the wave amplitudes in our 2D
models with this linear theory constitutes a diagnostic of whether our
2D solver is capable to properly solve for IGW propagation, at least for
waves that are indeed in the linear regime considered here.

\section{Results}
\label{sec:results}

\subsection{General observations on the dynamics}

\begin{figure}
\begin{center}
    \includegraphics[width=\columnwidth]{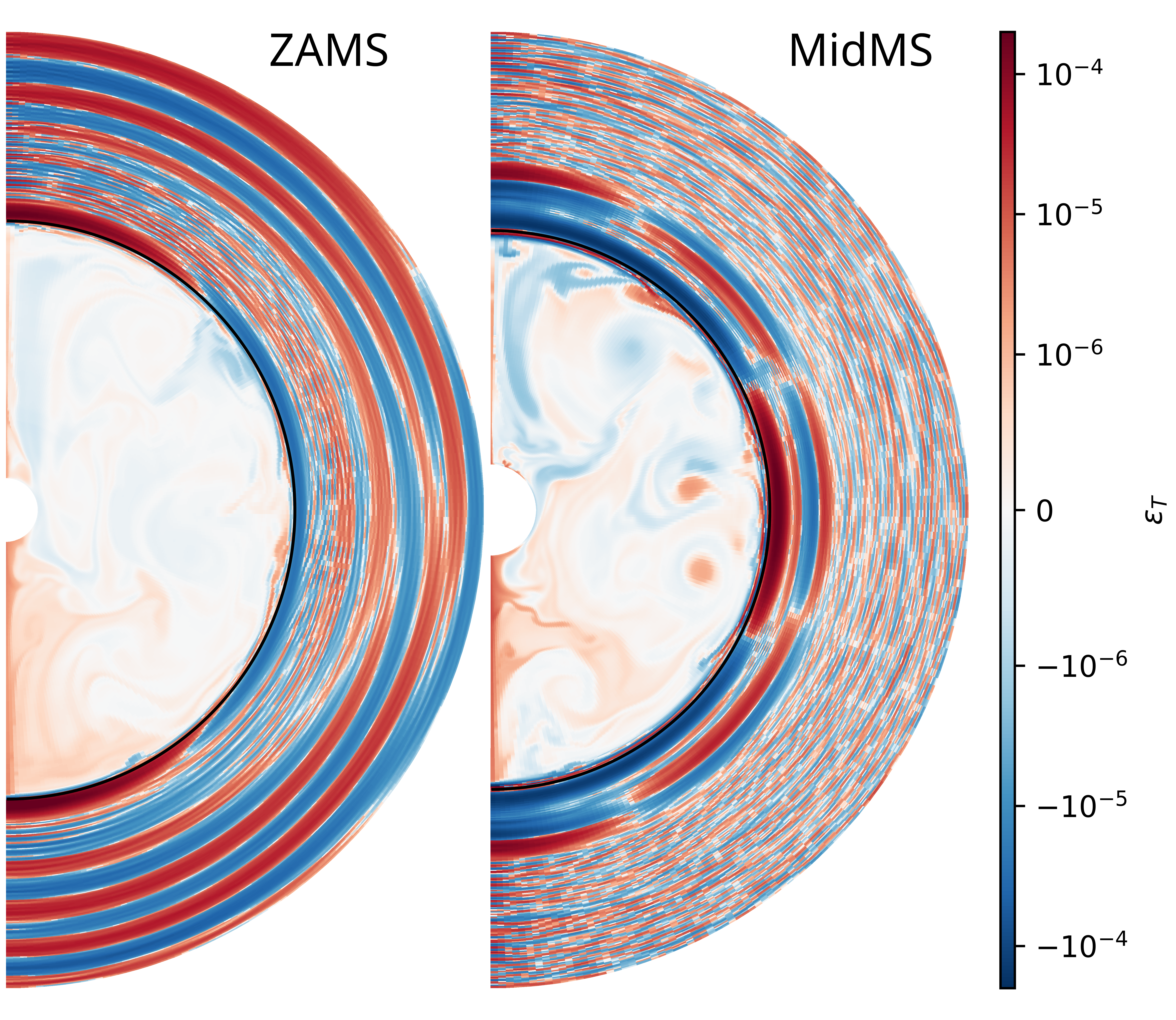}
\end{center}
    \caption{Temperature perturbation maps for the ZAMS (left) and MidMS
    (right) cases. The solid black line marks the position of the convective
    boundary.}
\label{fig:temppert}
\end{figure}

\Cref{fig:temppert} shows the instantaneous temperature perturbation
\begin{equation}
    \varepsilon_T = \frac{T - \lmean{T}_\theta}{\lmean{T}_\theta}
\end{equation}
for the ZAMS and MidMS cases. The TAMS case exhibit similar behavior to the
MidMS case. The convective part is extremely well mixed with $|\varepsilon_T|
\lesssim \num{e-6}$ while the stably stratified region shows stronger
deviations from the mean profile ($|\varepsilon_T| \sim \num{e-4}$) that organize in
well-defined patterns.  Those patterns are the signature of internal gravity
waves.  These plots already exhibit a striking difference between the ZAMS
model and evolved models: patterns of temperature perturbation have a similar angular extent
and amplitude throughout the stable layer in the ZAMS case, while two shells
with a different organization of $\varepsilon_T$ are clearly visible across the
stable region of the MidMS model. Sitting atop the core is the $N^2$-peak
region, showing patterns of $\varepsilon_T$ with larger angular and radial extent
than in the rest of the stable region at lower $N^2$.  \Cref{sub:waves} presents
an in-depth analysis of the wave content in those simulations.

\begin{figure}
\begin{center}
    \includegraphics[width=\columnwidth]{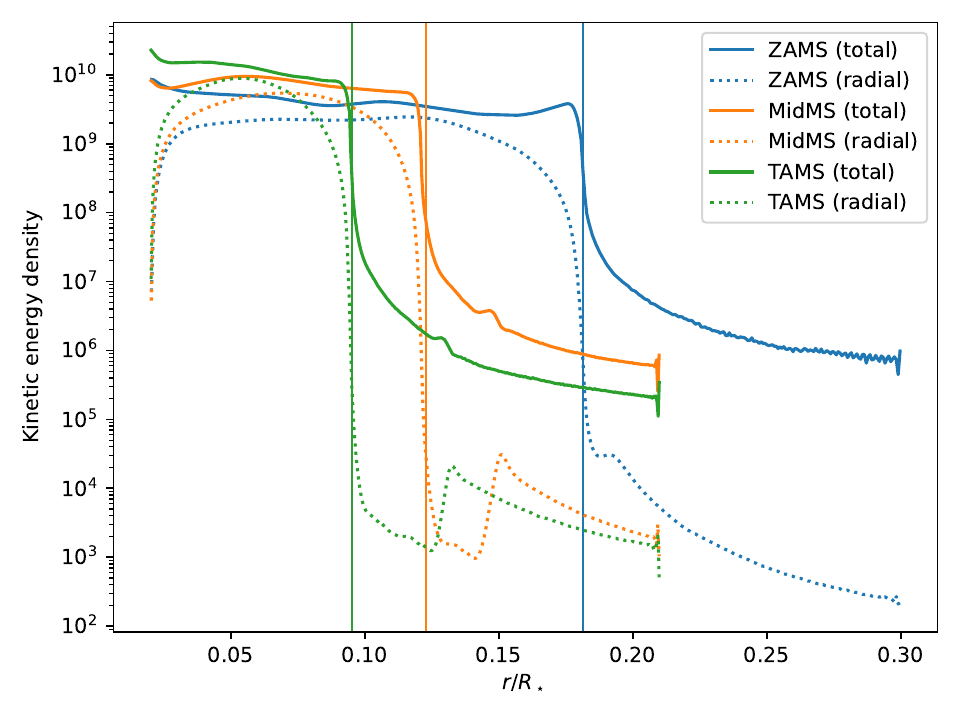}
\end{center}
    \caption{Kinetic energy density profiles for the three cases, total
    $\lmean{\lmean{\frac12\rho v^2}_\theta}_t$ and radial
    $\lmean{\lmean{\frac12\rho v_r^2}_\theta}_t$.
    The solid vertical lines mark the position of the convective core boundary
    according to the Schwarzschild criterion.}
\label{fig:kinprof}
\end{figure}

As can be seen on \Cref{fig:kinprof}, the kinetic energy density is constant in the
convective core, decreases sharply at the convective-core boundary, and
slightly decreases with radius in the stably stratified region for all three
models.  The kinetic energy density in the convective region is higher in the evolved
models compared to the ZAMS model by about a factor of three. This is expected
as the luminosity of the star (and therefore the energy feeding convection)
increases with age along the main sequence.  However, the drop of kinetic
energy at the convective boundary is much deeper for evolved models (4 orders
of magnitude) than for the
ZAMS model (about 3 orders of magnitude), resulting in an about 10-fold lower
kinetic energy density in the stable region of evolved models.  The $N^2$-peak layer of
evolved models reduces the transfer of kinetic energy
from the convective core to the radiative envelope.
\Cref{fig:kinprof} provides an
additional insight on the effects of the $N^2$-peak layer.  While the radial
component of the kinetic energy density is roughly monotonic in the ZAMS model and most
of the MidMS and TAMS radiative envelopes, it dips
by more than an order of magnitude in the $N^2$-peak layer of evolved models
compared to the rest of the stable region. This is readily explained by the
dispersion relation \cref{eq:dispersion} which implies that $v^2/v_r^2 \simeq
N^2/\omega^2$; as $N^2$ increases, the horizontal component of the kinetic
energy density will therefore carry more of the energy.

\subsection{Penetration length}
\label{sub:pendepthres}

This subsection focuses specifically on the transfer of
matter across the convective boundary via the estimation of an overshooting length
as described in \cref{sub:pendepth}.

Average values of the overshooting lengths $l_k^\text{ov}$ and
$l_{\delta T}^\text{ov}$ are shown in \Cref{tab:lmax} for the three models.
For the ZAMS case, one recovers the value found in
\cite{baraffe_study_2023} for the 5L0 case ($l_k\sim0.09 H_p$, $l_{\delta T}
\sim0.06 H_p$). For both evolved cases MidMS and TAMS, the penetration length
associated with either flux is much lower $l\sim 0.025 H_p$. It should be
noted that $l(\theta, t)$ is only one grid cell $\Delta r \sim 0.007 H_p$
in many areas, meaning the penetration length is likely under-resolved and
possibly over-estimated for those evolved cases. Overshooting is therefore
severely limited in the MidMS and TAMS models (likely due to the stabilizing
stratification in helium that builds up), and is unlikely
to be an efficient mixing and wave excitation mechanism above the convective
cores of massive stars which have evolved along the main sequence.

\begin{table}
    \begin{center}
        \begin{tabular}{*{5}c}
            Model & $l_k^\text{ov}$ (cm) & $l_{\delta T}^\text{ov}$ (cm) & $l_k^\text{ov}/H_{p,\text{conv}}$ & $l_{\delta T}^\text{ov}/H_{p,\text{conv}}$ \\
            \hline
            ZAMS & \num{1.53e9} & \num{1.51e9} & 0.084 & 0.083 \\
            MidMS & \num{3.91e8} & \num{4.86e8} & 0.022 & 0.028 \\
            TAMS & \num{3.65e8} & \num{4.64e8} & 0.022 & 0.028\\
        \end{tabular}
    \end{center}
    \caption{Overshooting length measured for the three evolution stages.}
    \label{tab:lmax}
\end{table}

\subsection{Wave study}
\label{sub:waves}

\begin{figure}
    \begin{center}
        \includegraphics[width=\columnwidth]{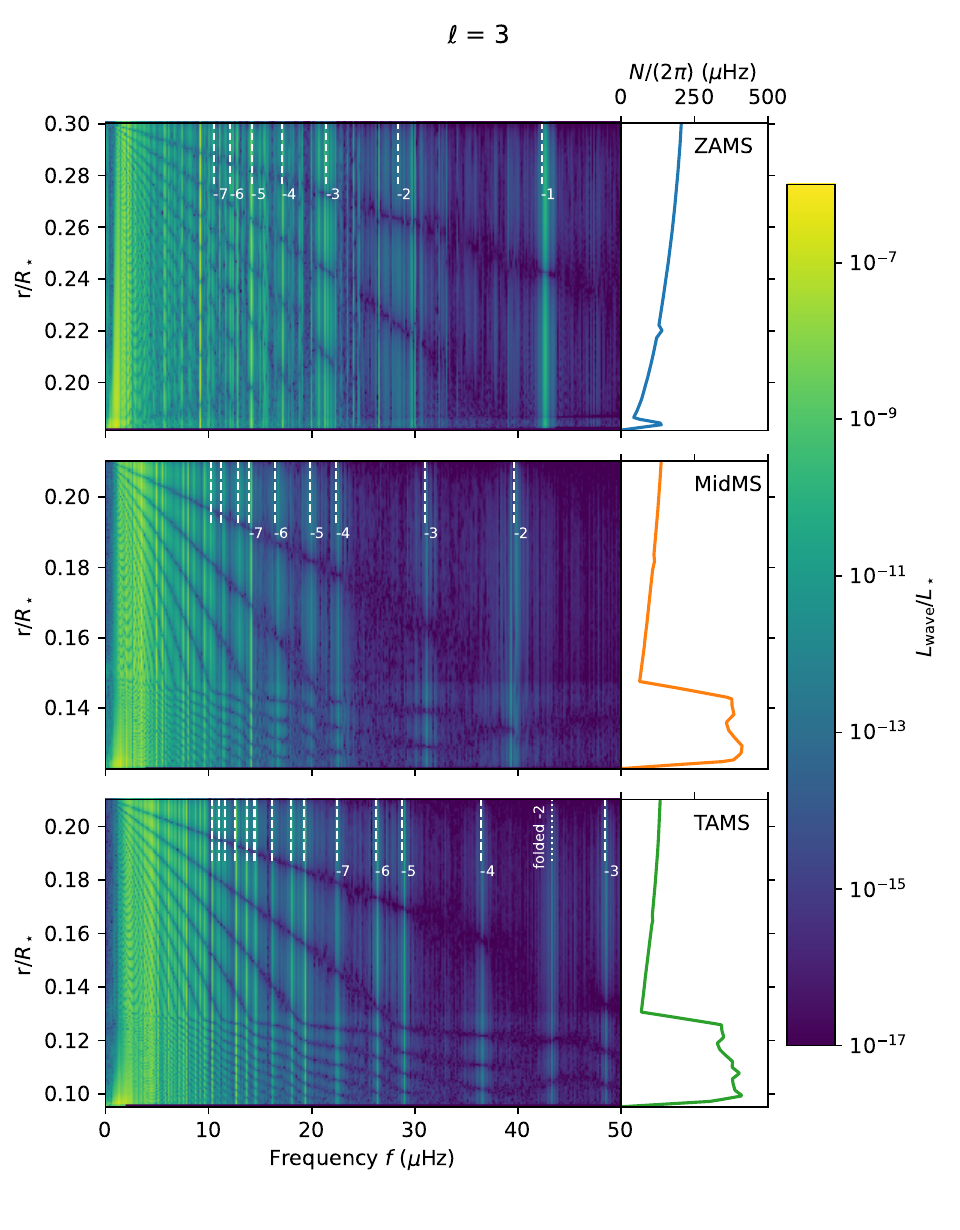}
    \end{center}
    \caption{Wave luminosity spectrum for all three models, along with the
    Brunt-Väisälä frequency profiles and g-modes. From top to bottom: ZAMS,
    MidMS, and TAMS. The dashed lines locate the g-modes predicted by GYRE
    (above \SI{10}{\micro\Hz}),
    with their radial order within the Eckart-Scuflaire-Osaki-Takata scheme
    \citep{takata_analysis_2006}. Note that the labels for modes -8 to
    -10 (MidMS) and modes -8 to -16 (TAMS) are omitted for readability purposes.
    The dotted line on the TAMS spectrum marks the position of an alias of
    the -2 mode (predicted at \SI{56.7}{\micro\Hz}) due to spectral folding
    around the Nyquist frequency \SI{50}{\micro\Hz}.
    The plot on the right is the Brunt-Väisälä frequency profile of
    each model, note that the frequency scale is different from the spectra.}
    \label{fig:wavespec}
\end{figure}

\begin{figure*}
\begin{center}
    \includegraphics[width=\textwidth]{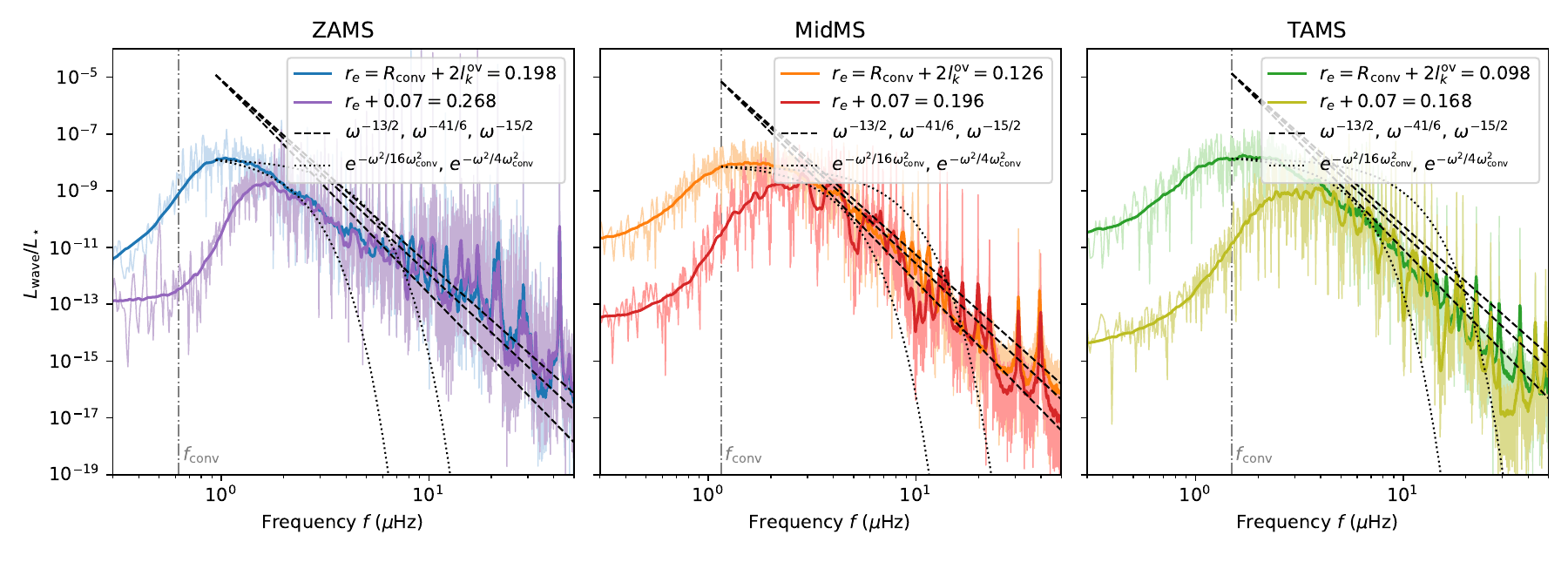}
\end{center}
\caption{Wave luminosity spectrum profiles at $\ell = 3$ for the three cases, at two
    different radii above the convective boundary. $r_e$ is slightly
    above the convective boundary (at twice the penetration length $l_k^\text{ov}$ found
    in \cref{sub:pendepthres}), and $r_e + 0.07\rstar$ is
    outside of the $N^2$-peak region of the MidMS and TAMS models.
    The thin light-colored solid lines are the spectra obtained from the
    simulations, and the thick dark-colored solid lines are the same profiles
    smoothed for readability purposes. The vertical gray
    dash-dotted line is the convective frequency. Dotted lines are gaussian
    models of excitation by plume penetration \citep{pincon_generation_2016},
    and dashed lines are scalings considering Reynolds stress
    \citep{lecoanet_internal_2013}.}
    \label{fig:waveprofs}
\end{figure*}

The previous subsection shows that the $N^2$-peak layer hinders radial motions
in the evolved models. This subsection focuses on the consequences on IGW
in the radiative region.
\Cref{fig:wavespec} shows the wave luminosity spectrum of all three models for $\ell=3$.
Vertical bright lines are the signature of g-modes, i.e. standing IGW that
form in the radiative zone. Dark lines follow the
radial nodes of the standing modes. As expected for IGW, the number
of nodes increases with decreasing $\omega$
\citep{aerts_asteroseismology_2010}. A notable feature of both evolved
models compared to the ZAMS model is that nodes are narrowed in the $N^2$-peak
layer in evolved models. This is, again, explained by the dispersion relation
\cref{eq:dispersion}: for a given frequency $\omega$, increasing the
Brunt-Väisälä frequency leads to a more horizontal wavefront.
We used the stellar oscillation code GYRE
\citep{townsend_gyre_2013} to compute the frequency of g-modes in the
simulated cavity. \Cref{fig:wavespec} shows a good match between the
predicted frequencies (white dashed lines) and the position of the peaks in the
power spectra, especially for the MidMS and TAMS model. The match is less
convincing for the ZAMS model at higher frequency, in particular for the $-2$
and $-3$ g-modes. This could be due to the stronger reflection that results in
noisier signal at high frequencies for this model (see \Cref{fig:waveprofs}).

\begin{figure*}
    \begin{center}
        \includegraphics[width=0.9\textwidth]{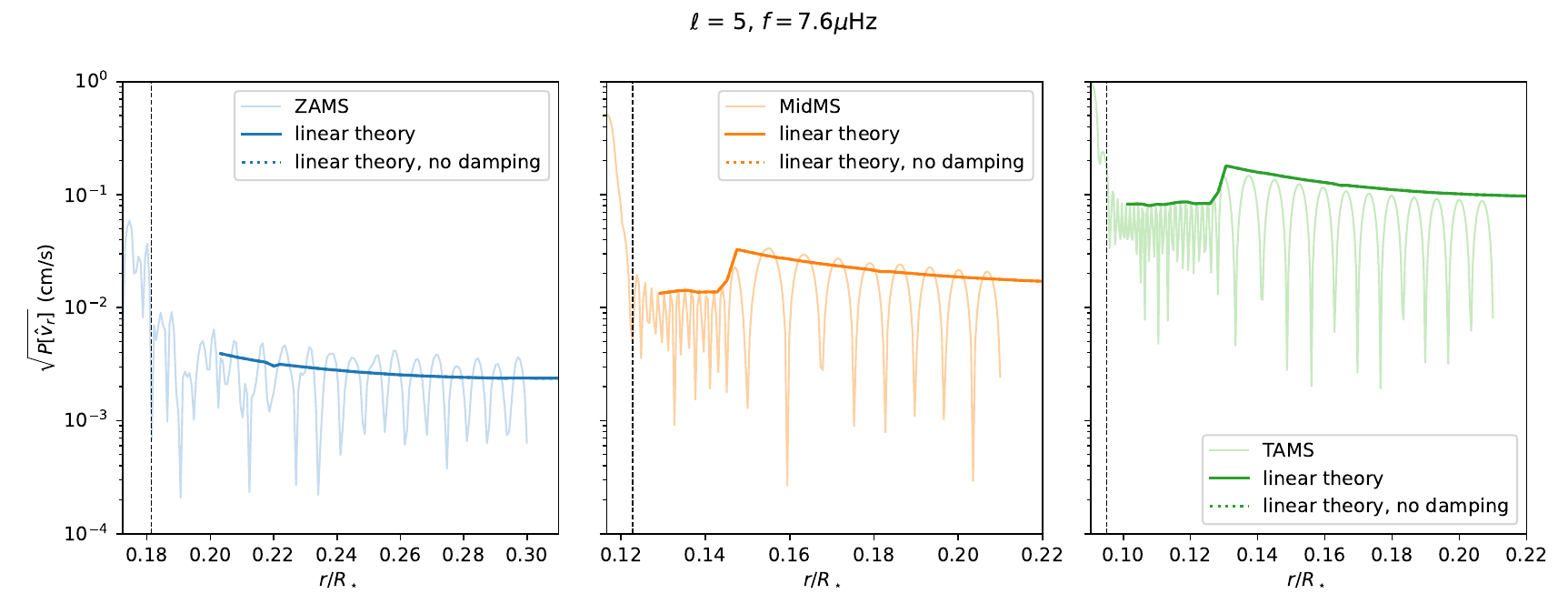}

        \includegraphics[width=0.9\textwidth]{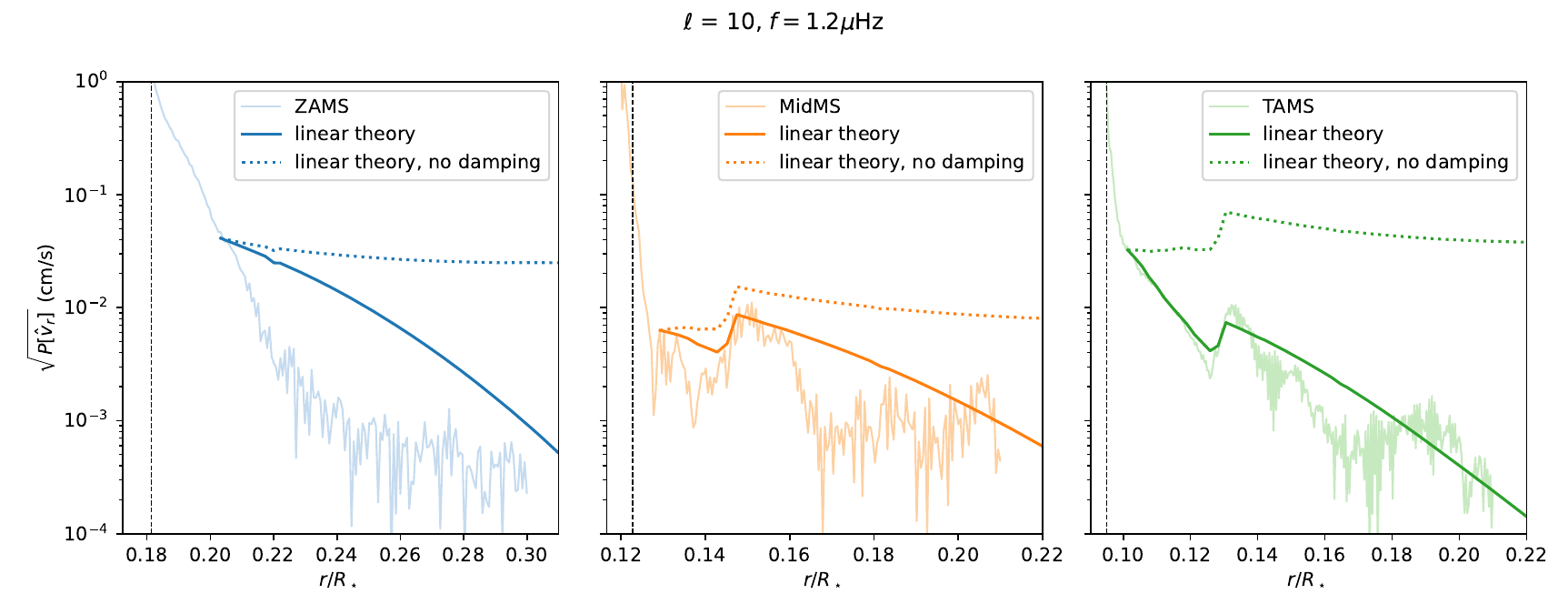}
    \end{center}
    \caption{Wave power spectrum and linear prediction for the three models, for
    two different modes. On each plot, the light solid line is the spectrum obtained
    from the simulation, the dark solid line is the linear prediction with
    radiative damping \cref{eq:linampl}, and the dotted line is the linear
    prediction without radiative damping. These two lines are superimposed
    for the $(l=5,f=\SI{7.6}{\micro\hertz})$ mode as damping is small.
    The black dashed line marks the convective boundary $R_\text{conv}$.}
    \label{fig:wavelin}
\end{figure*}

\Cref{fig:wavelin} compares power spectra at $(\ell=5,f=\SI{7.6}{\micro\Hz})$
and $(\ell=10,f=\SI{1.2}{\micro\Hz})$ with the linear theory
\cref{eq:linampl} for all three models, with damping following
\cref{eq:lindamping} and without damping ($\tau=0$). Note that since
we use realistic values for the heat conductivity $\chi$ (see
\cref{eq:rad_cond}), we expect our 2D models to exhibit realistic damping
profiles.
The $(\ell=5,f=\SI{7.6}{\micro\Hz})$ mode undergoes very limited damping
in the simulated region, and as can be seen on \Cref{fig:wavelin}, the
simulated spectrum matches well the linear theory for all three models.
On the other hand, the $(\ell=10,f=\SI{1.2}{\micro\Hz})$ mode is predicted
to be severely damped (which is a consequence of the higher $\ell$ and lower
$f$ compared to the previous mode). The match between the linear prediction
and the simulated spectrum is much less convincing for this mode. The ZAMS
model does not match the prediction. The MidMS model departs from the linear
prediction in many places, with slopes that are different from the linear
prediction everywhere except in a small region above the $N^2$-peak.
The TAMS model is the one that matches the most the linear prediction for this
mode, with a good match in the $N^2$-peak region and roughly following the
predicted slope in the rest of the stratified region.
The increase in amplitude when exiting the $N^2$-peak layer is systematically
higher in the simulations than what is predicted by the linear theory. This
linear theory is developed using the WKB approximation, and therefore assumes
that spatial derivatives of various parameters (such as $N^2$) are small; this
assumption is potentially not valid at the top of the $N^2$-peak region, and
the linear theory may thus incorrectly predict
the change of amplitude as waves exit that region.

The two modes discussed above were chosen as they are in different regimes: one is
only very weakly damped while the other undergoes strong damping. Similar
observations can be made on other modes: high-frequency modes are subject to
low damping and match well the theoretical prediction, while modes at low
frequency are more strongly damped and the simulated spectrum does not always
match the linear prediction. The poorer match of low-frequency modes could be
due to their higher amplitude than high-frequency modes (see
\Cref{fig:wavespec}): they are further away from the linear regime in which the
linear theory is valid. Moreover, low-frequency
IGW have a shorter radial wavelength and therefore are not as well resolved
radially, which could partly explain the poorer match for those modes.
Indeed, the radial wavelength $\lambda_r$ from the dispersion relation \cref{eq:dispersion}
is well above the cell width $\Delta r$ for the mode
$(\ell=5,f=\SI{7.6}{\micro\Hz})$ across all models
($6\Delta r < \lambda_r < 60\Delta r$), but the radial wavelength
for the mode $(\ell=10,f=\SI{1.2}{\micro\Hz})$ is comparable to $2\Delta r$ for
all three models and even smaller in the $N^2$-peak region: the latter mode is
therefore under-resolved.
It should finally be noted that the linear theory \cref{eq:linampl} concerns travelling waves,
while waves in the simulated domain are subject to reflection due to
the reflective boundary condition at the outer boundary, and to the
strong gradient of $N^2$ at the top of the $N^2$-peak region.
Reflection of waves in the overall domain and/or within the $N^2$-peak
region could explain the higher amplitude of some modes in the 2D
models compared to the linear prediction. Moreover, large oscillations
characteristic of g-modes make it difficult to compare to the linear
theory.

\Cref{fig:waveprofs} compares the wave power spectra at $\ell=3$ for the three
stellar models at two radii: at the excitation radius \cref{eq:re}, and $0.07\rstar$
above it (outside of the $N^2$-peak region in
evolved models). The damping at low $f\lesssim \SI{2}{\micro\Hz}$ is
clearly visible when comparing both
radial positions. The damping is stronger in the evolved models (three orders of
magnitude) compared to the ZAMS model (two orders of magnitude). The gaussian
shape aroud the peak of power spectra (around \SIrange{2}{5}{\micro\Hz}),
usually suggested to be related to the excitation of IGW by convective
penetration \citep[e.g.][]{pincon_generation_2016, edelmann_three-dimensional_2019,
le_saux_two-dimensional_2022}, is less clearly marked in the evolved models
than in the ZAMS model. This could be
a signature of the much lower penetration length in evolved models (see
\cref{sub:pendepthres}). The power spectra align along the slope predicted for
IGW excited by Reynolds stress at higher frequencies, although the spread in
amplitude due to high-amplitude g-modes in the simulated spectra makes it
impossible to discriminate which model is the most relevant in the present
case.

\Cref{fig:waveprofs} exhibits extremely marked peaks in wave luminosity
above \SI{10}{\micro\Hz} for the ZAMS model, that have much lower amplitude
in the more extended but otherwise identical simulation shown in
\citet{lesaux_two-dimensional_2023}. Those strong peaks are probably the
result of wave reflections in the truncated domain. The spectra for the
evolved case do not seem to be affected as much by reflection, which could
be a result of the overall lower kinetic energy in evolved models compared
to ZAMS (\Cref{fig:kinprof}).

\begin{figure}
\begin{center}
    \includegraphics[width=\columnwidth]{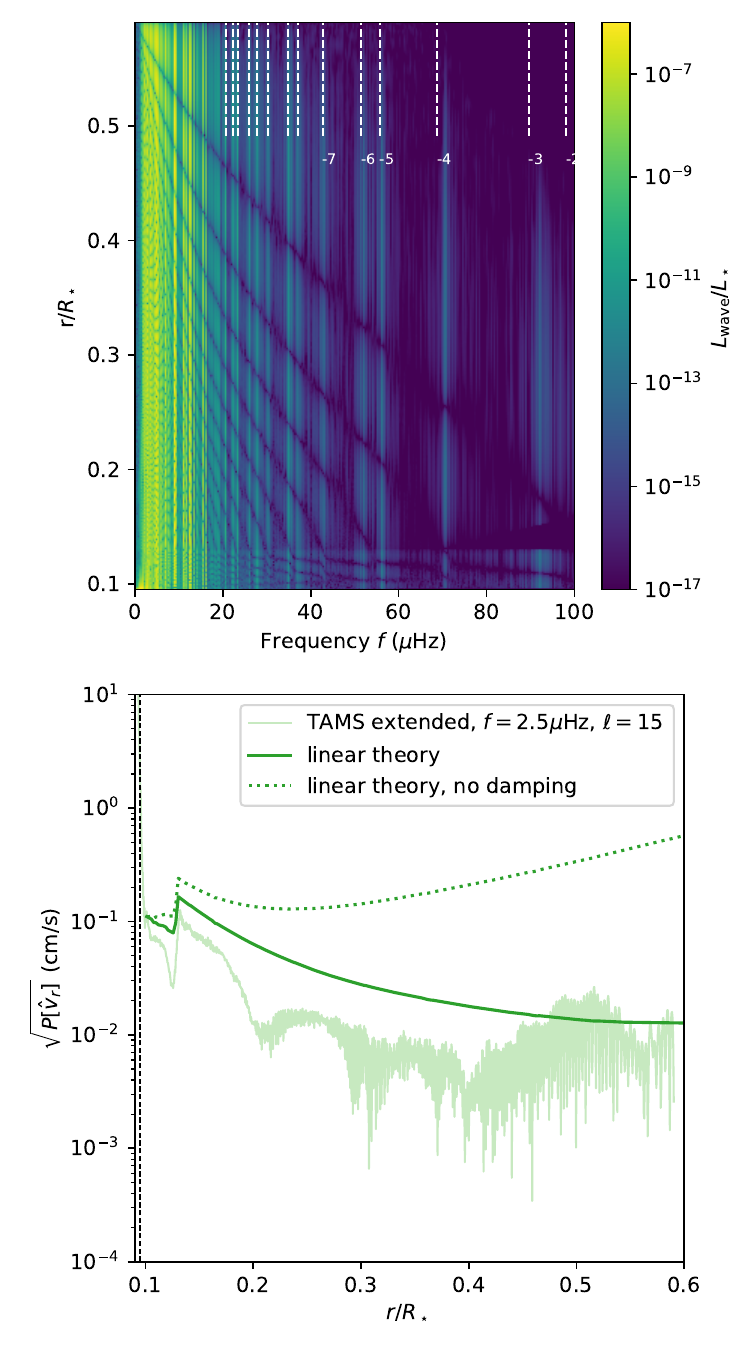}
\end{center}
    \caption{Wave luminosity spectrum for the extended TAMS model (top), and
    comparison with the linear theory for the
    $(\ell=15,f=\SI{2.5}{\micro\Hz})$ mode (bottom). Top: as in
    \Cref{fig:wavespec}, the dashed lines denote the g-modes predicted by GYRE.
    Labels for nodes -8 to -15 are omitted for readability. Bottom: as in
    \Cref{fig:wavelin}, the light solid line is the spectrum obtained from the
    simulation, the dark solid line is the linear prediction with radiative
    damping \cref{eq:linampl}, and the dotted line is the linear prediction
    without radiative damping.}
\label{fig:tams_extended}
\end{figure}

As a check, we performed an extended simulation of the TAMS
model (with $R_\text{out}=0.59\rstar$); \Cref{fig:tams_extended} shows
the obtained wave luminosity spectrum and comparison with the linear theory for
the $(\ell=15,f=\SI{2.5}{\micro\Hz})$ mode. The wave luminosity spectrum
shows similar features to the ones previously discussed for the truncated model. Note
that the g-modes are at different frequencies owing to the different aspect
ratio of the cavity in the extended case.
The simulated spectrum follows roughly
the slopes predicted by the linear theory for $r\lesssim0.2\rstar$ (with, again,
a stronger increase in amplitude than predicted when exiting the $N^2$-peak
layer at $r\simeq0.13\rstar$), but strong oscillations due to g-modes renders
the comparison to the theory difficult at higher radius.

\begin{figure}
    \begin{center}
        \includegraphics[width=0.45\textwidth]{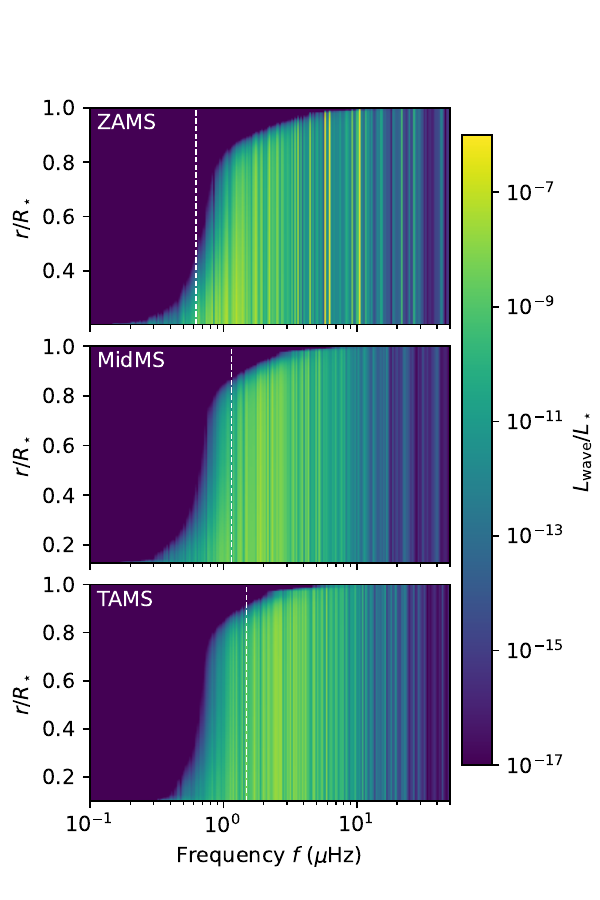}
    \end{center}
    \caption{Wave luminosity spectrum calculated from the linear theory
    \cref{eq:lwavelin} for all three models at $\ell=3$. The dashed line marks the
    convective frequency $f_{\text{conv}}$.}
    \label{fig:wavespecth}
\end{figure}

\Cref{fig:wavespecth} shows the theoretical wave power spectrum for the three
evolution stages, using the 1D initial profiles which extend up to $r=\rstar$.
This shows damping is very efficient at frequencies lower than about \SI{20}{\micro\Hz},
likely preventing most IGW from reaching the surface. IGW that are less damped
and susceptible of reaching the surface are at higher frequencies, but those
frequencies are weakly excited. Their predicted amplitude at the
surface is then extremely small. This confirms the results of
\citet{lesaux_two-dimensional_2023, anders_photometric_2023} for ZAMS models
and extends their conclusion to models evolved on the main sequence: IGW
excited by core convection likely have a very low amplitude at the surface.

\section{Conclusions}
\label{sec:conclusion}

The three simulations presented in this paper show several consequences of the
stratification in helium that builds up on top of the convective core of
massive stars during the evolution on the main sequence. The stratification in
helium is mechanically stable and strongly
limits the penetration of plumes in the radiative envelope. This hinders
mixing by overshooting in evolved stars compared to ZAMS stars.
Moreover, the convective frequency increases as stars evolve along the main
sequence, IGW excited by convection are then at higher frequency in older stars.
IGW propagation is limited in evolved stars due to several effects: damping in
the $N^2$-peak region, reflections at the top of the $N^2$-peak region
due to the steep change in stratification, and/or weaker excitation by plume
penetration. This reduces the overall power
of waves susceptible to reach the surface of an evolved star.
This strengthen the conclusions from \citet{lesaux_two-dimensional_2023,
anders_photometric_2023}. Our results clearly highlight that the amplitude of
IGW propagating in the radiative envelope of intermediate mass stars is
expected to be small. It is thus even
less likely to detect IGW excited by core convection in these stars, casting
doubts on the interpretation of \citet{bowman_photometric_2019,
bowman_photometric_2020}.

Finally, \citet{baraffe_study_2023} shows mixing by overshooting as calculated for ZAMS
models is already too weak to explain the observed width of the main
sequence in the Hertzsprung-Russell diagram for stellar masses $M\gtrsim 10\msun$.
Our study further strengthens this conclusion by showing that
overshooting only gets weaker compared to ZAMS as stars evolve along
the main sequence and a stable stratification in helium builds up above
the convective core.

Our results complement recent studies by \citet{vanon_three-dimensional_2023,
varghese_chemical_2023, ratnasingam_internal_2023}. They present simulations in
similar setups, studying the interaction between the convective core and
radiative zone of a massive star along the main sequence. It should be
underlined that those studies present 2D/3D simulations under the anelastic
approximation, without an helium mass fraction field, and with modified
heat fluxes, diffusivities and temperature profiles compared to stellar
evolution models to mimic the $N^2$-peak region above the core of evolved
models. Our simulations on the other hand are 2D, fully compressible, solving
for helium mass fraction conservation, and with diffusivities and temperature
profiles consistent with 1D stellar evolution models. In particular, including
a realistic helium mass fraction profile rather than mimicking the $N^2$-peak
via a modified temperature profile leads to a more realistic damping profile in
the $N^2$-peak region as we do not conflate $N_t$ and $N$ in
\cref{eq:lindamping}.

There are some differences between our findings and those
previous studies. \citet{ratnasingam_internal_2023, vanon_three-dimensional_2023}
find their power spectra have a better agreement with plume excitation, while
we find a better agreement with Reynolds stress excitation (\Cref{fig:waveprofs}).
\citet{vanon_three-dimensional_2023} report discrepancies between their TAMS
simulations and theoretical predictions of g-modes with GYRE. They
suggest some compressible effects that are not resolved by their
anelastic solver could be at play here, for example mixed modes. We however
find a good agreement between GYRE predictions and our TAMS models.
Finally, \citet{varghese_chemical_2023, vanon_three-dimensional_2023}
find that IGW become evanescent in radial domain for TAMS models, and suggest
that the WKB approximation does not apply in those models. This is in contrast
with our results, where we find a good agreement between the WKB linear theory
and our simulations (particularly for weakly-damped modes where this approximation
should be relevant), and do not observe the waves to become evanescent. The
reasons for all those discrepancies are unclear, and might stem from the
unrealistic temperature and diffusivity profiles used in these studies.
A comparison with the recent work by \citet{thompson_3d_2024} based on 3D
simulations of a $25\msun$ is also difficult since it does not include
radiative diffusion. As we show in present work and in
\citet{lesaux_two-dimensional_2023}, the propagation of low frequency waves in
the radiative zone is strongly impacted by radiative effects.

Our main findings are based on 2D numerical simulations and do not consider the
effect of rotation on convection and wave properties.
\citet{vanon_three-dimensional_2023} find similar power spectra in their 3D
simulations as the 2D ones from \citet{rogers_internal_2013}, suggesting the
power spectra we show in this 2D study should be largely unaffected by
dimensionality. Convective velocities in
2D simulations are usually larger than in 3D simulations
\citep{meakin_turbulent_2007, pratt_comparison_2020}, suggesting larger
overshooting lengths predicted by 2D simulations compared to 3D models. However, the
plume filling factor, which is linked to the convective flow structure, also
plays a role in the determination of an effective overshooting length and is
expected to be different between 2D and 3D simulations, with smaller filling
factors producing smaller overshooting lengths \citep{zahn_convective_1991,
brummell_penetration_2002, rogers_numerical_2006, pratt_comparison_2020}. How
the plume structure, which affects the determination of an overshooting length,
varies between 2D and 3D descriptions is still an open question.
One can however note that our results are qualitatively consistent with
\citet{vanon_three-dimensional_2023} who also report a decrease of overshooting
as stars evolve along the main sequence. We are
currently performing a detailed study of the filling factor and plume shape at
the convective boundary of stellar cores and envelopes using the same
simulation framework in 2D and 3D to address this question (Dethero et al.,
2024, in prep.). Regarding rotation, it can impact convective structures as
they align with the rotation axis, affecting both convective velocities and
wave excitation. As the Coriolis force deflects radial convective plumes at a
convective-radiative interface, one expects decreasing overshooting lengths as
the rotation rate increases \citep{brummell_penetration_2002,
rogers_internal_2013}. The effects of rotation and of the presence of a helium
stratification above the core for stars on the Main Sequence would thus combine
to limit convective penetration, making the problem found in ZAMS stellar stars
to reproduce observations \citep{baraffe_study_2023} even worse for rotating MS
stars. Regarding the impact of rotation on waves, recent 3D simulations by
\citet{anders_photometric_2023} of a $15\msun$ ZAMS model suggest that moderate
rotation rate can slightly boost the luminosity of waves excited at the
convective core boundary. This boosting effect on the wave luminosity may be
significant for fast rotators, but it will still be limited by the presence of
a strongly stratified layer above the convective core as the star evolves on
the main sequence. While the power excess observed at the surface of
intermediate mass stars and interpreted as due to internal waves excited by the
convective core \citep{bowman_photometric_2019} is ubiquitous, all these stars
are not fast rotators and not on the ZAMS.

To conclude, our work suggests that overshooting becomes weaker along the main
sequence as also found by \citet{vanon_three-dimensional_2023}. This aggravates
the problem pointed out in \citet{baraffe_study_2023} to reproduce observations
in the Hertzsprung-Russell diagram. In addition, the stabilizing helium
stratification that builds up above the core of massive stars on the main
sequence strongly affects the excitation and propagation of IGW. This limits
the amplitude of IGW reaching the surface in evolved models compared to ZAMS
models. Understanding IGW and overshooting signatures in the context of
long-term stellar evolution therefore requires further efforts on the modelling
and observational fronts.

\section*{Acknowledgements}
The authors thank an anonymous reviewer for their valuable comments which
helped clarify this paper.
This work is supported by the ERC grant No. 787361-COBOM
and the STFC Consolidated Grant ST/V000721/1.
The authors would like to acknowledge the use of the
University of Exeter HighPerformance Computing (HPC) facility ISCA and of the
DiRAC Data Intensive service at Leicester, operated by the University of
Leicester IT Services, which forms part of the STFC DiRAC HPC Facility. The
equipment was funded by BEIS capital funding via STFC capital grants
ST/K000373/1 and ST/R002363/1 and STFC DiRAC Operations grant ST/R001014/1.
DiRAC is part of the National e-Infrastructure. Part of this work was performed
under the auspices of the U.S. Department of Energy by Lawrence Livermore
National Laboratory under Contract DE-AC52-07NA27344 (LLNL-JRNL-859808).

\section*{Data availability}
Data backing this study will be shared on reasonable request to the
corresponding author.

\bibliographystyle{mnras}
\bibliography{references} 

\begin{thebibliography}{}
\makeatletter
\relax
\def\mn@urlcharsother{\let\do\@makeother \do\$\do\&\do\#\do\^\do\_\do\%\do\~}
\def\mn@doi{\begingroup\mn@urlcharsother \@ifnextchar [ {\mn@doi@}
  {\mn@doi@[]}}
\def\mn@doi@[#1]#2{\def\@tempa{#1}\ifx\@tempa\@empty \href
  {http://dx.doi.org/#2} {doi:#2}\else \href {http://dx.doi.org/#2} {#1}\fi
  \endgroup}
\def\mn@eprint#1#2{\mn@eprint@#1:#2::\@nil}
\def\mn@eprint@arXiv#1{\href {http://arxiv.org/abs/#1} {{\tt arXiv:#1}}}
\def\mn@eprint@dblp#1{\href {http://dblp.uni-trier.de/rec/bibtex/#1.xml}
  {dblp:#1}}
\def\mn@eprint@#1:#2:#3:#4\@nil{\def\@tempa {#1}\def\@tempb {#2}\def\@tempc
  {#3}\ifx \@tempc \@empty \let \@tempc \@tempb \let \@tempb \@tempa \fi \ifx
  \@tempb \@empty \def\@tempb {arXiv}\fi \@ifundefined
  {mn@eprint@\@tempb}{\@tempb:\@tempc}{\expandafter \expandafter \csname
  mn@eprint@\@tempb\endcsname \expandafter{\@tempc}}}

\bibitem[\protect\citeauthoryear{Aerts, Christensen-Dalsgaard  \& Kurtz}{Aerts
  et~al.}{2010}]{aerts_asteroseismology_2010}
Aerts C.,  Christensen-Dalsgaard J.,   Kurtz D.~W.,  2010, Asteroseismology.
Springer, \mn@doi{10.1007/978-1-4020-5803-5}

\bibitem[\protect\citeauthoryear{Anders et~al.,}{Anders
  et~al.}{2023}]{anders_photometric_2023}
Anders E.~H.,  et~al., 2023, \mn@doi [Nat Astron] {10.1038/s41550-023-02040-7},
  7, 1228

\bibitem[\protect\citeauthoryear{Baraffe \& El~Eid}{Baraffe \&
  El~Eid}{1991}]{baraffe_evolution_1991}
Baraffe I.,  El~Eid M.~F.,  1991, A\&A, 245, 548

\bibitem[\protect\citeauthoryear{Baraffe, Chabrier, Allard  \&
  Hauschildt}{Baraffe et~al.}{1998}]{baraffe_evolutionary_1998}
Baraffe I.,  Chabrier G.,  Allard F.,   Hauschildt P.~H.,  1998, \mn@doi [A\&A]
  {10.48550/arXiv.astro-ph/9805009}, 337, 403

\bibitem[\protect\citeauthoryear{Baraffe, Pratt, Vlaykov, Guillet, Goffrey,
  Le~Saux  \& Constantino}{Baraffe et~al.}{2021}]{baraffe_two-dimensional_2021}
Baraffe I.,  Pratt J.,  Vlaykov D.~G.,  Guillet T.,  Goffrey T.,  Le~Saux A.,
  Constantino T.,  2021, \mn@doi [A\&A] {10.1051/0004-6361/202140441}, 654,
  A126

\bibitem[\protect\citeauthoryear{Baraffe et~al.,}{Baraffe
  et~al.}{2023}]{baraffe_study_2023}
Baraffe I.,  et~al., 2023, \mn@doi [MNRAS] {10.1093/mnras/stad009}, 519, 5333

\bibitem[\protect\citeauthoryear{Bossini et~al.,}{Bossini
  et~al.}{2015}]{bossini_uncertainties_2015}
Bossini D.,  et~al., 2015, \mn@doi [MNRAS] {10.1093/mnras/stv1738}, 453, 2290

\bibitem[\protect\citeauthoryear{Bowman et~al.,}{Bowman
  et~al.}{2019}]{bowman_photometric_2019}
Bowman D.~M.,  et~al., 2019, \mn@doi [A\&A] {10.1051/0004-6361/201833662}, 621,
  A135

\bibitem[\protect\citeauthoryear{Bowman, Burssens, Simón-Díaz, Edelmann,
  Rogers, Horst, Röpke  \& Aerts}{Bowman
  et~al.}{2020}]{bowman_photometric_2020}
Bowman D.~M.,  Burssens S.,  Simón-Díaz S.,  Edelmann P. V.~F.,  Rogers
  T.~M.,  Horst L.,  Röpke F.~K.,   Aerts C.,  2020, \mn@doi [A\&A]
  {10.1051/0004-6361/202038224}, 640, A36

\bibitem[\protect\citeauthoryear{Brummell, Clune  \& Toomre}{Brummell
  et~al.}{2002}]{brummell_penetration_2002}
Brummell N.~H.,  Clune T.~L.,   Toomre J.,  2002, \mn@doi [ApJ]
  {10.1086/339626}, 570, 825

\bibitem[\protect\citeauthoryear{Claret \& Torres}{Claret \&
  Torres}{2016}]{claret_dependence_2016}
Claret A.,  Torres G.,  2016, \mn@doi [A\&A] {10.1051/0004-6361/201628779},
  592, A15

\bibitem[\protect\citeauthoryear{Edelmann, Ratnasingam, Pedersen, Bowman, Prat
  \& Rogers}{Edelmann et~al.}{2019}]{edelmann_three-dimensional_2019}
Edelmann P. V.~F.,  Ratnasingam R.~P.,  Pedersen M.~G.,  Bowman D.~M.,  Prat
  V.,   Rogers T.~M.,  2019, \mn@doi [ApJ] {10.3847/1538-4357/ab12df}, 876, 4

\bibitem[\protect\citeauthoryear{Goffrey et~al.,}{Goffrey
  et~al.}{2017}]{goffrey_benchmarking_2017}
Goffrey T.,  et~al., 2017, \mn@doi [A\&A] {10.1051/0004-6361/201628960}, 600,
  A7

\bibitem[\protect\citeauthoryear{Iglesias \& Rogers}{Iglesias \&
  Rogers}{1996}]{iglesias_updated_1996}
Iglesias C.~A.,  Rogers F.~J.,  1996, \mn@doi [ApJ] {10.1086/177381}, 464, 943

\bibitem[\protect\citeauthoryear{Le~Saux et~al.,}{Le~Saux
  et~al.}{2022}]{le_saux_two-dimensional_2022}
Le~Saux A.,  et~al., 2022, \mn@doi [A\&A] {10.1051/0004-6361/202142569}, 660,
  A51

\bibitem[\protect\citeauthoryear{Le~Saux, Baraffe, Guillet, Vlaykov, Morison,
  Pratt, Constantino  \& Goffrey}{Le~Saux
  et~al.}{2023}]{lesaux_two-dimensional_2023}
Le~Saux A.,  Baraffe I.,  Guillet T.,  Vlaykov D.~G.,  Morison A.,  Pratt J.,
  Constantino T.,   Goffrey T.,  2023, \mn@doi [MNRAS]
  {10.1093/mnras/stad1067}, 522, 2835

\bibitem[\protect\citeauthoryear{Lecoanet \& Quataert}{Lecoanet \&
  Quataert}{2013}]{lecoanet_internal_2013}
Lecoanet D.,  Quataert E.,  2013, \mn@doi [MNRAS] {10.1093/mnras/stt055}, 430,
  2363

\bibitem[\protect\citeauthoryear{Lecoanet et~al.,}{Lecoanet
  et~al.}{2019}]{lecoanet_low-frequency_2019}
Lecoanet D.,  et~al., 2019, \mn@doi [ApJ] {10.3847/2041-8213/ab5446}, 886, L15

\bibitem[\protect\citeauthoryear{Lecoanet, Cantiello, Anders, Quataert,
  Couston, Bouffard, Favier  \& Le~Bars}{Lecoanet
  et~al.}{2021}]{lecoanet_surface_2021}
Lecoanet D.,  Cantiello M.,  Anders E.~H.,  Quataert E.,  Couston L.-A.,
  Bouffard M.,  Favier B.,   Le~Bars M.,  2021, \mn@doi [MNRAS]
  {10.1093/mnras/stab2524}, 508, 132

\bibitem[\protect\citeauthoryear{Meakin \& Arnett}{Meakin \&
  Arnett}{2007}]{meakin_turbulent_2007}
Meakin C.~A.,  Arnett D.,  2007, \mn@doi [ApJ] {10.1086/520318}, 667, 448

\bibitem[\protect\citeauthoryear{Miglio, Montalbán, Noels  \&
  Eggenberger}{Miglio et~al.}{2008}]{miglio_probing_2008}
Miglio A.,  Montalbán J.,  Noels A.,   Eggenberger P.,  2008, \mn@doi [MNRAS]
  {10.1111/j.1365-2966.2008.13112.x}, 386, 1487

\bibitem[\protect\citeauthoryear{Pinçon, Belkacem  \& Goupil}{Pinçon
  et~al.}{2016}]{pincon_generation_2016}
Pinçon C.,  Belkacem K.,   Goupil M.~J.,  2016, \mn@doi [A\&A]
  {10.1051/0004-6361/201527663}, 588, A122

\bibitem[\protect\citeauthoryear{Pratt, Baraffe, Goffrey, Constantino, Viallet,
  Popov, Walder  \& Folini}{Pratt et~al.}{2017}]{pratt_extreme_2017}
Pratt J.,  Baraffe I.,  Goffrey T.,  Constantino T.,  Viallet M.,  Popov M.~V.,
   Walder R.,   Folini D.,  2017, \mn@doi [A\&A] {10.1051/0004-6361/201630362},
  604, A125

\bibitem[\protect\citeauthoryear{Pratt, Baraffe, Goffrey, Geroux, Constantino,
  Folini  \& Walder}{Pratt et~al.}{2020}]{pratt_comparison_2020}
Pratt J.,  Baraffe I.,  Goffrey T.,  Geroux C.,  Constantino T.,  Folini D.,
  Walder R.,  2020, \mn@doi [A\&A] {10.1051/0004-6361/201834736}, 638, A15

\bibitem[\protect\citeauthoryear{Press}{Press}{1981}]{press_radiative_1981}
Press W.~H.,  1981, \mn@doi [ApJ] {10.1086/158809}, 245, 286

\bibitem[\protect\citeauthoryear{Ratnasingam, Rogers, Chowdhury, Handler,
  Vanon, Varghese  \& Edelmann}{Ratnasingam
  et~al.}{2023}]{ratnasingam_internal_2023}
Ratnasingam R.~P.,  Rogers T.~M.,  Chowdhury S.,  Handler G.,  Vanon R.,
  Varghese A.,   Edelmann P. V.~F.,  2023, \mn@doi [A\&A]
  {10.1051/0004-6361/202245727}

\bibitem[\protect\citeauthoryear{Rogers \& Nayfonov}{Rogers \&
  Nayfonov}{2002}]{rogers_updated_2002}
Rogers F.~J.,  Nayfonov A.,  2002, \mn@doi [ApJ] {10.1086/341894}, 576, 1064

\bibitem[\protect\citeauthoryear{Rogers, Glatzmaier  \& Jones}{Rogers
  et~al.}{2006}]{rogers_numerical_2006}
Rogers T.~M.,  Glatzmaier G.~A.,   Jones C.~A.,  2006, \mn@doi [ApJ]
  {10.1086/508482}, 653, 765

\bibitem[\protect\citeauthoryear{Rogers, Lin, McElwaine  \& Lau}{Rogers
  et~al.}{2013}]{rogers_internal_2013}
Rogers T.~M.,  Lin D. N.~C.,  McElwaine J.~N.,   Lau H. H.~B.,  2013, \mn@doi
  [ApJ] {10.1088/0004-637X/772/1/21}, 772, 21

\bibitem[\protect\citeauthoryear{Rosenfield et~al.,}{Rosenfield
  et~al.}{2017}]{rosenfield_new_2017}
Rosenfield P.,  et~al., 2017, \mn@doi [ApJ] {10.3847/1538-4357/aa70a2}, 841, 69

\bibitem[\protect\citeauthoryear{Schatzman}{Schatzman}{1993}]{schatzman_transport_1993}
Schatzman E.,  1993, A\&A, 279, 431

\bibitem[\protect\citeauthoryear{Shaviv \& Salpeter}{Shaviv \&
  Salpeter}{1973}]{shaviv_convective_1973}
Shaviv G.,  Salpeter E.~E.,  1973, \mn@doi [ApJ] {10.1086/152318}, 184, 191

\bibitem[\protect\citeauthoryear{Stancliffe, Fossati, Passy  \&
  Schneider}{Stancliffe et~al.}{2016}]{stancliffe_confronting_2016}
Stancliffe R.~J.,  Fossati L.,  Passy J.-C.,   Schneider F. R.~N.,  2016,
  \mn@doi [A\&A] {10.1051/0004-6361/201527099}, 586, A119

\bibitem[\protect\citeauthoryear{Takata}{Takata}{2006}]{takata_analysis_2006}
Takata M.,  2006, \mn@doi [Publ. Astron. Soc. Jpn.] {10.1093/pasj/58.5.893},
  58, 893

\bibitem[\protect\citeauthoryear{Thompson, Herwig, Woodward, Mao, Denissenkov,
  Bowman  \& Blouin}{Thompson et~al.}{2024}]{thompson_3d_2024}
Thompson W.,  Herwig F.,  Woodward P.~R.,  Mao H.,  Denissenkov P.,  Bowman
  D.~M.,   Blouin S.,  2024, \mn@doi [MNRAS] {10.1093/mnras/stae1162}, 531,
  1316

\bibitem[\protect\citeauthoryear{Townsend \& Teitler}{Townsend \&
  Teitler}{2013}]{townsend_gyre_2013}
Townsend R. H.~D.,  Teitler S.~A.,  2013, \mn@doi [MNRAS]
  {10.1093/mnras/stt1533}, 435, 3406

\bibitem[\protect\citeauthoryear{Vanon, Edelmann, Ratnasingam, Varghese  \&
  Rogers}{Vanon et~al.}{2023}]{vanon_three-dimensional_2023}
Vanon R.,  Edelmann P. V.~F.,  Ratnasingam R.~P.,  Varghese A.,   Rogers T.~M.,
   2023, \mn@doi [ApJ] {10.3847/1538-4357/ace9db}, 954, 171

\bibitem[\protect\citeauthoryear{Varghese, Ratnasingam, Vanon, Edelmann  \&
  Rogers}{Varghese et~al.}{2023}]{varghese_chemical_2023}
Varghese A.,  Ratnasingam R.~P.,  Vanon R.,  Edelmann P. V.~F.,   Rogers T.~M.,
   2023, \mn@doi [ApJ] {10.3847/1538-4357/aca092}, 942, 53

\bibitem[\protect\citeauthoryear{Viallet, Goffrey, Baraffe, Folini, Geroux,
  Popov, Pratt  \& Walder}{Viallet et~al.}{2016}]{viallet_jacobian-free_2016}
Viallet M.,  Goffrey T.,  Baraffe I.,  Folini D.,  Geroux C.,  Popov M.~V.,
  Pratt J.,   Walder R.,  2016, \mn@doi [A\&A] {10.1051/0004-6361/201527339},
  586, A153

\bibitem[\protect\citeauthoryear{Zahn}{Zahn}{1991}]{zahn_convective_1991}
Zahn J.~P.,  1991, A\&A, 252, 179

\bibitem[\protect\citeauthoryear{Zahn}{Zahn}{1992}]{zahn_circulation_1992}
Zahn J.~P.,  1992, A\&A, 265, 115

\bibitem[\protect\citeauthoryear{Zahn, Talon  \& Matias}{Zahn
  et~al.}{1997}]{zahn_angular_1997}
Zahn J.-P.,  Talon S.,   Matias J.,  1997, A\&A, 322, 320

\makeatother
\end{thebibliography}

\bsp	
\label{lastpage}
\end{document}